\documentclass[12pt]{article}
\usepackage[a4paper,margin=1in]{geometry}

\usepackage{amsmath}
\usepackage{graphicx,psfrag,epsf}
\usepackage{enumerate}
\usepackage[numbers]{natbib}
\usepackage{url} 
\usepackage{hyperref}
\usepackage{tablefootnote}
\usepackage{threeparttable}
\usepackage{graphicx}  
\usepackage{tabularx}
\usepackage{tabulary}
\usepackage{float}
\usepackage{multirow}
\usepackage{comment}
\usepackage{amsmath,amssymb,amsthm}
\usepackage{graphicx}
\usepackage{hyperref}
\usepackage{setspace}
\usepackage{geometry}
\usepackage{tocloft}
\usepackage{natbib}
\usepackage{booktabs} 
\usepackage{siunitx}

\usepackage{caption}
\usepackage{setspace}
\newcommand{\blind}{0}

\begin{document}




\if0\blind
{
  \title{\bf  Bayesian analysis of the causal reference-based model for missing data in clinical trials accommodating partially observed post-intercurrent event data}
  \author{Brendah Nansereko \\
    Department of Medical Statistics, London School of Hygiene and Tropical Medicine \\
     \\
    Marcel Wolbers \\
    Data and Statistical Sciences, Pharma Development, Roche, Basel, Switzerland \\
     \\    
    James R. Carpenter \\
    Department of Medical Statistics, London School of Hygiene and Tropical Medicine \\
         \\
    Jonathan W. Bartlett \\
    Department of Medical Statistics, London School of Hygiene and Tropical Medicine \\}
  \maketitle
} \fi

\if1\blind
{
  \bigskip
  \bigskip
  \bigskip
  \begin{center}
    {\LARGE\bf Title}
\end{center}
  \medskip
} \fi

\bigskip

\begin{abstract}
When treatment policy estimands are of interest, clinical trials often attempt to collect patient data after intercurrent events (ICEs), although such data are often limited. Retrieved dropout imputation methods, which use pre-ICE and available post-ICE data to impute missing post-ICE outcomes, are commonly applied but often yield treatment effect estimates with large standard errors (SEs) and may encounter convergence issues when post-ICE data are sparse. Reference-based imputation methods are also used, but they rely on strong assumptions about post-ICE outcomes, which can lead to biased estimates if these assumptions are incorrect.

To address these limitations, we previously proposed the reference-based Bayesian causal model (BCM), which incorporates a prior on the maintained effect parameter to reflect uncertainty in reference-based assumptions for missing post-ICE data. Our earlier work assumed no post-ICE data were observed. Here, we extend the BCM to incorporate available post-ICE outcomes, providing an approach that mitigates limitations of both retrieved-dropout and standard reference-based methods. We propose both a fully Bayesian model and an imputation-based approach.

A simulation study was conducted to evaluate the frequentist properties of the proposed methods in settings with partially observed post-ICE data and to compare performance with existing approaches. Retrieved-dropout methods produced higher estimated SEs than the BCM, particularly when post-ICE data were sparse. Under the BCM, treatment effect SEs increased as post-ICE data became more limited for both modelling approaches. Importantly, this increase can be controlled through the prior variance of the maintained effect parameter, with more informative priors stabilising estimation when post-ICE data are scarce.

\end{abstract}

\noindent%
\vfill

\newpage
\section{Introduction}

The ICH E9(R1) addendum provided a framework for the construction of estimands, which aids in the precise description of treatment effects in randomised controlled trials (RCTs). It comprises five key components , which integrate both clinical considerations and so-called intercurrent events (ICEs) \cite{RN5}. These components are the treatment conditions of interest, the patient target population, the end point, the population-level summary measure, and how ICEs are to be handled. The ICH E9(R1) guideline provides a number of strategies for integrating ICEs into the estimand definition. These include the treatment policy strategy, the hypothetical strategy, and other approaches that ensure a structured and consistent framework for estimand specification in clinical trials.

Following the ICH E9 guidelines, it is typical for clinical trials to attempt to collect patient outcome data after the occurrence of intercurrent events (ICE). Aligned with the intent-to-treat (ITT) principle, the treatment policy strategy seeks to assess the treatment effect including the effects of ICEs on subsequent outcomes. Data after the occurrence of the ICE are used directly in the estimation of the treatment effect under the treatment policy strategy. Following an ICE, two scenarios may arise. The patient can continue to be observed, allowing the collection of post-ICE data; these are sometimes referred to as retrieved dropout data (RD). Alternatively, the patient may become lost to follow-up immediately or at a later stage, resulting in missing data. In many studies, despite the best efforts of trial investigators, some post-ICE data may be missing, posing challenges for estimation and analysis. For example, discontinuation of randomised treatment can substantially increase the risk of missing subsequent clinic visits even when trial investigators attempt to continue visits for such patients.

Several methods have been proposed to handle missing data in the presence of ICEs. Methods such as the Mixed Model for Repeated Measures (MMRM) and standard multiple imputation (MI) under a missing at random (MAR) assumption ignore the occurrence of the ICEs and the missing data is (implicitly or explicitly) imputed under MAR. For missing post-ICE data, standard MAR-based methods (effectively) assume the conditional distribution of missing (post-ICE) data is the same as the corresponding conditional distribution in those who did not have the ICE if there is no or little post-ICE data observed data in the trial. This assumption is often unrealistic in RCTs, given that the occurrence of an ICE typically implies some change in the patient's treatment or condition.
Given that ICH E9(R1) mandates alignment between the statistical methods used for handling missing data and the predefined strategy for addressing ICEs within the estimand framework, standard MAR-based approaches are typically not appropriate in the context of estimating treatment policy effects \cite{RN14}.

The reference-based imputation (RBI) methods proposed by Carpenter \textit{et al.} assume that the distribution of missing (post-ICE) data for patients who experience an ICE in a given treatment arm is based (in a specified way) on the corresponding outcome distribution of the reference arm \cite{RN14}. 
RBI methods are primarily relevant to treatment discontinuation following ICEs in placebo-controlled trials, but may also be useful for other types of ICEs. The original proposal for RBI was made in the context of a trial where outcome data are not available post-ICE. When some post-ICE data are observed, in some software implementations (e.g. the R package rbmi \cite{RN78}), observed post-ICE data are not used in the imputation model fitting process, but when calculating the imputation distribution, the rbmi package conditions on all observed data, including any observed post-ICE values \cite{RN78}. Consequently, the imputed missing post-ICE values are impacted by observed post-ICE data, if present.
A limitation of current reference-based methods is their reliance on strong assumptions about post-ICE outcomes, which, if incorrect, will generally lead to biased estimates of the treatment effect. In addition, the imputation model fit does not utilise any observed post-ICE data to learn about the effect of the ICE on outcomes.

An alternative set of approaches, often termed `RD methods', impute missing data based on pre-ICE data and available post-ICE data \cite{RN58}. A range of MI implementations have been proposed and investigated that adjust in some way for the ICE occurrence information, and that rely on a different MAR assumption to the MMRM and `standard MI' mentioned earlier. Wang and Hu proposed a retrieved-dropout method, which imputes the missing post-ICE data at the final time point using information from the individuals who experience the ICE \cite{RN70}.  MMRM methods have also been expanded to accommodate the post-ICE data through the inclusion of the ICE variable \cite{RN69}. However, these methods often lead to treatment effect estimates with large standard errors, especially when the amount of retrieved data after the ICE is limited \cite{RN44}. Furthermore, the retrieved dropout models are prone to convergence failure, again  due to the limited amount of available post-ICE data \cite{RN77}.

 Cro \textit{et al.} recently proposed a retrieved dropout reference-based method for imputing missing post-ICE data \cite{RN76}. This method imputes the missing data by combining both the parameters obtained from a core reference-based model and the retrieved dropout model parameters. Under this method, the traditional reference-based model is extended to include offset parameters $(\gamma_{akj}^*)$ which capture departures from the reference-based assumption and whose values are informed by observed post-ICE outcomes. The $\gamma_{akj}^*$ parameters are estimated using a Bayesian model fitted to the retrieved dropout data, with priors specified.
A potential drawback of this approach is that it requires the specification of a multivariate prior for the $\gamma_{akj}^*$ parameters at different visits, which can be challenging due to the large number of parameters involved, particularly when there are many follow-up visits. Furthermore, estimation of these parameters may be difficult when no post-ICE data are observed for some ICE patterns defined by the timing of the ICE, especially when non-informative priors are used.
Given the above limitations with the available approaches, there is a need for methods that can utilise the post-ICE data when available, can handle the common situation when such post-ICE data is limited, and appropriately acknowledge uncertainty about the missing data assumptions.

White \textit{et al} proposed the causal reference-based imputation model, which uses potential outcomes and explicitly defines assumptions about how the occurrence of an ICE affects subsequent outcomes \cite{RN36}. In our previous work, we used this causal model and proposed the Reference-Based Bayesian Causal Model (BCM), which introduces a prior on the magnitude of the treatment effect that is maintained after the ICE occurs \cite{RN86}. In our previous work, we considered the setting where no post-ICE data were available and developed a Bayesian inference method. In practice, it is increasingly common for some post-ICE data to be observed in clinical trials. These observations should be utilised, both to improve the precision of the estimated treatment effect and because they provide valuable information about the distribution of the missing data.
As such, in this paper, we propose an extension of the BCM approach to  incorporate post-ICE observations. We also develop imputation-based approaches based on the BCM, which rely on the reference-based causal model only insofar as it is needed to handle imputation of missing data. For concreteness of exposition, until the discussion, we focus on a setting with a single type of ICE—treatment discontinuation, although the proposed methods could potentially be applied to other types of ICEs.

In Section 2, we describe the original BCM and its extension that incorporates post-ICE observations. This section also includes the description of imputation approaches based on the BCM. Section 3 describes a simulation study conducted to evaluate the performance of the proposed method alongside existing approaches. In Section 4, we apply the BCM methods to an antidepressant dataset with some post-ICE data available, and in Section 5, we conclude with a discussion.

\section{Reference-based Bayesian causal model}

White \textit{et al.} introduced a reference-based causal imputation model for handling missing post-ICE data within the potential outcomes framework, which we now describe \cite{RN36}.
Their setup is motivated for a setting with discontinuation of randomised treatment as the ICE.
Let \( Y_j(s) \) denote the potential outcome at visit \( j \) for a given patient if, possibly contrary to fact, they receive the active treatment for the first \( s \) visits, followed by the control or reference treatment for the remaining time.  Define \( D \) as the last visit before the occurrence of an ICE for an individual patient. Define \( Y(s) \) as the vector of such potential outcomes across all \( j_{\text{max}} \) follow-up visits under this treatment scenario. The sub-vectors \( Y_{\leq j}(s) \) and \( Y_{>j}(s) \) refer to the potential outcomes at and before visit \( j \), and strictly after visit \( j \), respectively. The expected values of these vectors are denoted by: $\mu(s) = {E}[Y(s)]$, $\mu_{\leq j}(s) = {E}[Y_{\leq j}(s)]$, $\mu_{>j}(s) = {E}[Y_{>j}(s)]$. The variance-covariance matrix of \( Y(s) \) is given by $\Sigma(s) = \text{Var}(Y(s))$. The corresponding submatrices are: $\Sigma_{\leq j \leq j}(s) = \text{Var}(Y_{\leq j}(s))$, $\Sigma_{> j \leq j}(s) = \text{Cov}(Y_{>j}(s), Y_{\leq j}(s))$, $\Sigma_{\leq j > j}(s) = \Sigma_{> j \leq j}(s)^\top$, $\Sigma_{>j >j}(s) = \text{Var}(Y_{>j}(s))$. The regression coefficients linking past to future outcomes are defined as:
\[
\beta_j(s) = \Sigma_{> j \leq j}(s) \, \Sigma_{\leq j \leq j}(s)^{-1}.
\]
For example, \( \beta_j(j) \) represents the multivariate regression coefficients for predicting future potential outcomes \( Y_{> j}(j) \) from prior outcomes \( Y_{\leq j}(j) \), assuming the patient received active treatment up to visit \( j \). 

Actual (as opposed to counterfactual) missing outcomes in the control arm are assumed to be MAR. We assume an MMRM for the full-data model of the hypothetical untreated/control outcomes across all visits \(Y(j_{\max})\).

\begin{align*}
Y_{j}(0) &= \mu_{ j}(0) + X_i^\top \alpha_{j} + \epsilon_{i  j}, \quad \varepsilon_i = (\varepsilon_{i1},\dots,\varepsilon_{ij_{max}})^{\top} \sim N(0,\,\Sigma(0)), 
\end{align*}

where \( X_i \) is a vector of baseline covariates, \( \alpha_{j}\) is the vector of regression coefficients for baseline covariate effects on outcome at visit $j$. We also assume an MMRM for the full-data model of the hypothetical fully treated outcomes across all visits \(Y(j_{\max})\). This model is specified as:

\begin{align*}
Y_{j}(j_{max}) &= \mu_{ j}(j_{max}) + X_i^\top \alpha_{j} + \epsilon_{i  j}, \quad \varepsilon_i = (\varepsilon_{i1},\dots,\varepsilon_{ij_{max}})^{\top} \sim N(0,\,\Sigma(j_{max})), 
\end{align*}

The central assumption of the reference-based causal model is that mean outcomes at the post-ICE visits can be expressed as a function of the difference in mean outcomes between treatment arms (i.e. the treatment effects) at the pre-ICE visits through a matrix valued maintained effect parameter $K_j$. In the potential outcome notation defined earlier, this is that
\begin{equation*}
E[Y_{>j}(j) - Y_{>j}(0) ] = K_jE[Y_{\leq j}(j) - Y_{\leq j}(0) ]
\end{equation*} 

where \(K_j\) is a \((j_{\max} - j) \times (j + 1)\) matrix of parameters for the maintained treatment effect.



White \textit{et al.}  proposed a simpler single-parameter model for the difference in mean outcomes at visit $u$ after discontinuation at visit $j$:
\begin{equation}
	E[Y_{u}(j) -  Y_{u}(0) ] =k_0E[Y_{j}(j) - Y_{j}(0)] 
\end{equation}

\begin{figure}[h]
  \centering
  \includegraphics[width=\textwidth]{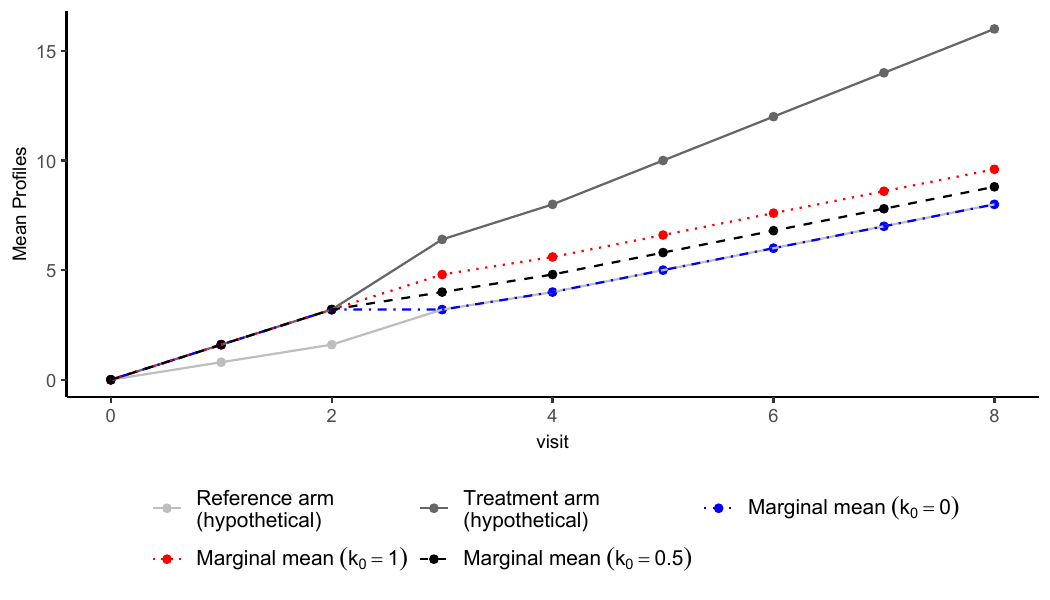}
 	\caption[Implied mean trajectories under the causal model.]{Implied mean trajectories under the causal model. The blue, red and black lines show the implied mean trajectories for someone who discontinues active treatment after visit 2  for $k_0=0$, $k_0=0.5$, and $k_0=1$, respectively. \label{fig:1_paper2}}
 \end{figure}

Under this single-parameter model, $k_0$ is a real-valued maintained effect parameter that reflects the user’s assumptions about how the treatment effect is maintained or decays after discontinuation of active treatment, as illustrated in Figure \ref{fig:1_paper2}. Under this specification, the maintained treatment effect is assumed to remain constant across all post-ICE visits. 

 Under White's single-parameter causal model, the post-ICE outcomes conditional on the pre-ICE outcomes are multivariate normal, with the conditional mean of the post-ICE outcomes \( Y_{>t} \) given by:
\begin{equation}
	\label{eq:1_paper2}
	{E}(Y_{>j} \mid Y_{\leq j}, T = a, D = j) = \beta_j(j) Y_{\leq j} - \beta_j(j) \mu_{\leq j}(j) + k_0(\mu_j(j) - \mu_j(0)) + \mu_{>j}(0),
\end{equation}

The residual covariance matrix for post-ICE outcomes \( Y_{>j} \) given the pre-ICE outcomes $Y_{\leq j}$ is given by \[\Omega_j(j)\] 
where
\[ \quad 
\Omega_j(s) = \Sigma_{>j,<j}(s)\,\Sigma_{<j,<j}(s)^{-1}\,\Sigma_{>j,<j}(s)^{\top}.
\]

A number of the originally proposed RBI variants are special cases of the causal model, with a particular choice of the parameter $k_0$:
\begin{itemize}
    \item $k_0 = 1$: all of the treatment effect is maintained after the ICE, corresponding to the copy increments in reference (CIR) assumption.
    \item $k_0 = 0$: none of the treatment effect is maintained after the ICE, corresponding to the jump to reference (J2R) assumption.
\end{itemize}

We assume that the variance--covariance matrices are identical across all discontinuation times, 
i.e., \(\Sigma(s) \equiv \Sigma\).
This assumption is consistent with the MMRM, where a common variance--covariance structure across treatment arms is frequently adopted.  It is plausible in settings where the ICE, such as treatment discontinuation, is not expected to alter the variability or correlation structure of the outcomes.

In our earlier work, we extended this approach by proposing the Bayesian causal model (BCM), which builds upon the reference-based causal imputation framework introduced by White \textit{et al} \cite{RN86}.  Our model incorporates a prior distribution on the `maintained effect' parameter $k_0$, thereby relaxing the strong assumptions inherent in traditional reference-based methods, which assume the value of $k_0$ to be known. This method is implemented within a Bayesian framework. In our previous work, we assumed that no post-ICE data were available.

White \textit{et al}'s paper supplementary appendix D shows the derivation of an explicit expression for the treatment policy treatment effect at the final time point based on the causal model \cite{RN36}. This equation is expressed in terms of the hypothetical on-treatment means, the maintained effect parameter $k_0$ and the proportion of ICEs occurring at each visit in the active arm. Given draws from the posterior distributions of the model parameters, this can be used to obtain an estimate of the posterior mean of the treatment effect at the final visit. Let $\hat{\theta}^l$ denote the $l$-th posterior draw ($l = 1, \dots, L$) for a generic parameter $\theta$. The treatment effect at the final time point under the BCM is estimated as:


\begin{equation}
	\label{eq:2_paper2}
	\begin{aligned}
		\hat \theta_{BCM} = \frac{1}{L} \sum_{l=1}^L \Big[ &\hat \pi^l_{j_{\text{max}}} \left( \hat \mu^l_{j_{\text{max}}}(j_{\text{max}}) - \hat \mu^l_{j_{\text{max}}}(0) \right) + \\
		&\sum_{j < j_{\text{max}}} \hat \pi^l_j \hat k_0^l \left( \hat \mu^l_{j}(j) - \hat \mu^l_{j}(0) \right) \Big]
	\end{aligned}
\end{equation}

where $\hat{\pi}_j$ denotes the proportions of patients in the active treatment arm who discontinue at visit $j$, for $j = 1, \dots, j_{\text{max}}$.
This estimator is a combination of: (i) $\hat{\mu}$ values, which are posterior draws of the means from the MMRM; (ii) posterior draws of the maintained treatment effect parameter $\hat{k}_0$; and (iii) posterior draws of the proportions $\hat{\pi}_j$ which are the proportions of patients in the active arm who discontinue at visit $j$.

\subsection{Bayesian causal model with post-ICE data}

We now extend the BCM to incorporate any available post-ICE data. This is implemented under the Bayesian framework using the \texttt{Stan} software. The change required relative to the implementation when no post-ICE data are observed is to include the additional likelihood contributions corresponding to the observed post-ICE data, conditional on the pre-ICE data. These likelihood contributions for the post-ICE data correspond precisely to the imputation distribution in the causal model developed by White \textit{et al.}. For simplicity, here and in our simulations we assume either that a participant is either missing all post-ICE visit outcomes or has them all observed. As such, for participants who discontinue treatment at time \( j \) but who have their subsequent outcomes observed, the likelihood contribution corresponding to the multivariate normal distribution of $Y_{>j}|Y_{\leq j}, D=j$ is included, with mean as given in Equation \ref{eq:1_paper2}.
As in the initial BCM with no post-ICE data, the treatment effect at the final time point is estimated using Equation \ref{eq:2_paper2}, as derived by White \textit{et al}. 

In our earlier work, where no post-ICE data were assumed to be observed, the posterior for the $k_0$ parameter remained unchanged from the prior, since there were no data to inform its estimation. However, with some post-ICE data observed, the data are informative about the value of $k_0$, and so the model can learn about its value from the data as well as the prior. When no post-ICE data are available, the value of $k_0$ must be fixed (assumed known) or an informative prior used. When some post-ICE data are observed, we can potentially use a weaker or even essentially flat prior for $k_0$. We discuss prior choice further in Section \ref{sec:priorchoice}.

\subsection{Imputation approach for the Bayesian causal model}

In this subsection, we propose two imputation-based approaches which utilise the Bayesian causal model. Our motivation for using imputation rather than direct Bayes as described in the previous subsection is the idea that we may only want to use the Bayesian causal model so far as it is needed to handle the missing data. Specifically, if we do have complete or almost complete post-ICE data, we would want our inference to be essentially the same as the complete data analysis inference (e.g. based on a simple ANCOVA of the final time point outcome). Generally, if we use the direct/full Bayes approach, this will not be the case, whereas it is if we use imputation. An imputation approach may moreover be particularly advantageous in settings with multiple ICEs, where different assumptions may apply to different events. Unlike the initial BCM approach, which integrates imputation and analysis within a Bayesian framework, an imputation approach separates the two, offering greater flexibility.

We propose two approaches for imputing post-ICE data under the causal model: a conditional mean imputation method and a multiple imputation–based method. 
The multiple imputation approach is combined with the bootstrap. We use the bootstrap variance rather than Rubin's rules-based variances because the imputation and analysis models are uncongenial under Rubin's rules, leading to an upward bias in the Rubin's MI variance estimate relative to the repeated--sampling variance. 
For each bootstrap sample, the parameters of the causal model are estimated by the  maximum a posteriori values (MAP) using the \texttt{Stan} package's built-in optimisation routines (specifically, the optimising() function), which avoids the need for MCMC sampling. These parameter estimates are then used to construct the multivariate normal distribution from which post-ICE outcomes are imputed, generating \(M\) complete datasets. Each dataset is analysed separately using a regression of the final time point outcome on baseline characteristics and treatment arm. Within each bootstrap sample, \(M\) imputations are created. The treatment effect parameter within each bootstrap sample is calculated as the average of the \(M\) estimates obtained from the imputed datasets, and the overall treatment effect is obtained using the original dataset using the same procedure. The standard errors are estimated from the empirical variability of the bootstrap-sample mean estimates (i.e. the sampling distribution of the mean treatment effect across bootstrap samples).

In the conditional mean approach, the missing post-ICE values are imputed using the conditional expectations derived from the causal model \cite{RN40}. The parameters of the causal model are estimated via MAP optimisation using the \texttt{Stan} package and substituted into Equation~\ref{eq:1_paper2} to obtain the conditional means for post-ICE outcomes in the active treatment arm. 
The missing data in the reference arm are also imputed using the MAP estimates under the MAR assumption. Wolbers \textit{et al.} showed that conditional mean imputation, a deterministic method, yields similar estimates to the Bayesian MI for the standard reference-based imputation approach \cite{RN40}. Wolbers \textit{et al.} also showed that the deterministic conditional mean imputation approach is equivalent to the MI approach described above, with an infinite number of random imputations, provided that the analysis model is a linear model such as ANCOVA. Standard errors under the BCM conditional mean imputation approach can be estimated using the jackknife or the bootstrap method. The jackknife has the advantage of avoiding resampling variability, in contrast to the bootstrap, and thus yields a deterministic procedure when used with conditional mean imputation.

\subsection{Choice of priors}\label{sec:priorchoice}

The BCM requires specification of prior distributions for the parameters in Equation~\ref{eq:1_paper2}, including the hypothetical mean parameters \(\mu_j\), the proportions of patients experiencing an intercurrent event (ICE) at each visit in the active treatment arm \(\pi_j\), and the maintained treatment-effect parameter \(k_0\). Non-informative normal priors may be specified for the \(\mu_j\) parameters, while a Dirichlet prior is a natural choice for the vector of proportions \(\pi_j\). 

The parameter \(k_0\) captures the extent to which the treatment effect is maintained or decays following the ICE, and its prior specification should reflect the analyst’s assumptions about the post-ICE outcome distribution. \(k_0 > 0\)  means that if there was a benefit of active treatment before the ICE, there is still some benefit afterwards, whereas  \(k_0 < 0\)  means that a benefit of the active treatment before the ICE leads to a worse outcome (compared to control) post-ICE.

A normal prior may be specified for $k_0 \sim N(\mu_{k_0},\sigma^2_{k_0})$, allowing for either a maintained beneficial effect $(k_0>0)$ or a switch from benefit to harmful effect $(k_0<0)$. Alternatively, priors that constrain $k_0$ to be strictly positive may be used when there is strong substantive prior belief-for example, a strictly positive prior can be specified if it can be assumed that a benefit of active treatment while it is taken could not translate into worse outcomes (on average) after discontinuation. Values of $k_0$ outside the interval $[0,1]$ could be deemed implausible in settings where the ICE corresponds to treatment discontinuation; hence, priors such as a Beta distribution may be appropriate.

When a substantial amount of post-ICE data is observed, weakly informative or flat priors may be specified for $k_0$, and inference under the fully Bayesian approach will be driven primarily by the observed data rather than the prior. In this setting, the MAP estimates obtained under the BCM imputation approach will be close to maximum likelihood estimates. However, when post-ICE data are sparse, which is often the case in practice, more informative priors are required. In such settings, the resulting estimates and inferences will be influenced by the prior specification under both the fully Bayesian approach and the BCM multiple imputation approach.

\section{Simulation Study}

We conducted a simulation study to investigate the frequentist properties of the proposed BCM and imputation approaches when applied to settings with some available post-ICE data. In addition, we compared the performance of the BCM-based methods with that of a retrieved dropout method and traditional RBI. The objectives of the simulation study were therefore twofold: (i) to assess bias, precision, and frequentist coverage of credible/confidence intervals of the proposed causal model approaches under different scenarios of ICE rates and missingness; and (ii) to evaluate how the performance of the new methods compares to traditional RBI and retrieved dropout methods.

\subsection{Simulation Scenarios}
The design of our simulation study was adapted from Bell et al.~\cite{RN77}, which in turn was modelled to reflect the structure of the PIONEER 1 trial~\cite{RN56}. This trial evaluated the efficacy of oral semaglutide monotherapy versus placebo in patients with type 2 diabetes. In PIONEER 1, HbA1c levels were measured at weeks 0, 4, 8, 14, 20, and 26 for both treatment and placebo arms.

To mirror this setting, we simulated hypothetical on-treatment outcomes separately for the treatment and placebo groups, generating outcomes for 500 patients per arm from a multivariate normal distribution:

\[
Y_{iT} \sim MVN(\mu_T, \Sigma)
\]

where $T=a$ and $T=r$ correspond to the active treatment and control arms, respectively. The parameters used for the simulation are summarized in Table~\ref{1_paper2}.

\begin{table}[H]
	\centering
	\begin{threeparttable}
		\caption{Simulation parameters for HbA1c outcomes at each visit (\(t_j\)) over 26 weeks}
		\label{1_paper2}
		\begin{tabularx}{\textwidth}{X|X|X|X} \hline
			Visit (\(t_j\)) & Mean (\(\mu_a\)) & Mean (\(\mu_p\)) & Variance* (\(\Sigma\)) \\ \hline
			0    & 7.92  & 7.92 & 0.48 \\
			4    & 7.55  & 7.82 & 0.80 \\
			8    & 7.20  & 7.80 & 1.10 \\
			14   & 7.10  & 7.80 & 1.40 \\
			20   & 7.05  & 7.78 & 1.23 \\
			26   & 7.05  & 7.78 & 1.48 \\ \hline
		\end{tabularx}
		\begin{tablenotes}
			\item[*] The values represent the diagonal elements of the covariance matrix. The covariance matrix was constructed using a first-order spatial power structure: \(\text{cor}(t_i, t_j) = \rho^{|t_i - t_j| / 4}\), with \(\rho = 0.8\).
		\end{tablenotes}
	\end{threeparttable}
\end{table}

Treatment discontinuation was modeled under two intercurrent event (ICE) scenarios: low (25\% in the control arm and 15\% in the active treatment arm) and high (60\% in the control arm and 50\% in the active treatment arm) discontinuation rates. At each post-baseline visit \(j\), the probability of discontinuation was simulated under a logistic regression, incorporating baseline values, prior outcomes, and treatment assignment:

\[
\text{logit}(P(D_i = j \mid D_i \geq j, Y_{i0}, Y_{ij-1}, T)) = \beta_{0}^m + \beta_{base}^T \times Y_{i0} + \beta_{prev}^T \times Y_{ij-1}
\]

Here, \(\beta_{base}^T\) and \(\beta_{prev}^T\) are treatment-specific regression coefficients (with \(T = a, p\) for active and placebo), selected based on Bell et al.~\cite{RN77}. The intercept \(\beta_0^m\) (\(m = low, high\)) was calibrated to achieve the target overall discontinuation rates in both treatment arms.

Off-treatment outcomes for patients in the active group who discontinued were simulated using the causal imputation model (Equation~\ref{eq:1_paper2}). We considered two data-generating scenarios, with true values $k_0=0$ and $k_0=1$ corresponding to J2R and CIR, respectively. For patients in the control group who discontinued treatment, the distribution of off-treatment outcomes was assumed to remain unchanged after the ICE.

Missing post-ICE data were generated under a missing completely at random (MCAR) mechanism among those who had the ICE. For each individual who experienced an ICE, we randomly determined whether they had complete post-ICE outcomes or whether all post-ICE outcomes were missing. We considered four scenarios defined by two scenarios of treatment discontinuation (low and high) crossed with two probabilities of missingness. The resulting missing data scenarios following treatment discontinuation are summarised in Table~\ref{3_paper2}. 

\begin{table}[H]
	\centering
	\begin{threeparttable}
		\caption{Logistic regression parameters for discontinuation probability}
		\label{2_paper2}
		\begin{tabularx}{\textwidth}{X|X|X|X|X|X|X} \hline
			Week & \(\beta_{base}^A\) & \(\beta_{prev}^A\) & \(\beta_{base}^P\) & \(\beta_{prev}^P\) & \(\beta_0^{low}\) & \(\beta_0^{high}\) \\ \hline
			8    & 0.30 & 1.14 & 0.30 & 1.14 & -15 & -13 \\
			14   & 0.10 & 1.47 & 0.10 & 1.33 & -15 & -13 \\
			20   & 0.05 & 1.48 & 0.05 & 1.51 & -15 & -13 \\
			26   & 0.00 & 1.40 & 0.00 & 1.46 & -15 & -13 \\ \hline
		\end{tabularx}
	\end{threeparttable}
\end{table}

\begin{table}[htbp]
    \centering
    \begin{threeparttable}
        \caption{Treatment discontinuation and missingness scenarios}
        \label{3_paper2}
        \begin{tabularx}{\textwidth}{p{8cm}|X|X|X|X} \hline
            \multicolumn{5}{c}{\textbf{Scenarios (\%)}} \\ \hline
              & LD--LM & LD--HM & HD--LM & HD--HM \\ \hline
            \textbf{Active treatment}  & & & &  \\
            Overall probability of discontinuation & 15 & 15 & 50 & 50 \\
            Probability of missingness among those with ICE     & 20 & 90 & 20 & 90 \\[2pt]
            \textbf{Control treatment} & & & &  \\
            Overall probability of discontinuation & 25 & 25 & 60 & 60 \\
            Probability of missingness among those with ICE     & 20 & 90 & 20 & 90 \\ \hline
        \end{tabularx}
    
        \begin{tablenotes}
        \scriptsize
        \item[] LD = low discontinuation; HD = high discontinuation; LM = low missingness; HM = high missingness
        \end{tablenotes}
        \end{threeparttable}
\end{table}

\subsection{Estimators}

For each of 5{,}000 simulated datasets, we applied the following estimators of treatment effect at the final visit:

\begin{itemize}
   \item \textbf{Complete-data analysis:} 
    We conducted a complete-data analysis by simulating a fully observed dataset in which post-ICE outcomes were generated under the causal model using the true parameter values. An ANCOVA model was then used to estimate the treatment effect at the final time point, assuming that all post-ICE data were available. Because, in practice, it is often not feasible to collect complete post-ICE data, this complete-data scenario serves as a benchmark against which we compare the methods described in this section for handling missing post-ICE outcomes.
    
    We also analyzed the fully observed dataset with a Bayesian BCM with a prior of $k_0 \sim N(0,100)$. This analysis was performed to compare the resulting estimates with those obtained from the complete-data ANCOVA model.
    
    \item \textbf{Retrieved-dropout imputation:} This method uses parameters estimated from the available post-ICE data to impute missing outcomes. The imputation model incorporates observed data from previous visits together with a parameter that captures each patient's discontinuation pattern by that visit. This approach is similar to the Pattern Intercepts Common Slopes (PICS) model described by Drury~\cite{RN58}, which includes separate intercept terms for each treatment discontinuation pattern up to the ICE timepoint~$D=j$. Retrieved-dropout imputation was performed using sequential imputation in the \texttt{mice} package in R. Both patients who experienced an ICE and those who did not contributed to the estimation of the imputation model, which was specified as a normal linear regression model, separate for each time point. 
    At each time point \( j \), the imputation model included a covariate representing the discontinuation pattern at that visit, defined as the time since discontinuation and coded as 0 for patients who had not yet discontinued by visit \( j \), and treated as a numeric variable thereafter. The model additionally included the baseline and previously observed outcomes \( Y_{<j} \). The imputation model for the outcome at visit \( j \) is expressed as:

    \[
    Y_{j} = \lambda_0 + \lambda T + \alpha\, P_{j} + \sum_{k=0}^{j-1} \lambda_{k+1} Y_k + \varepsilon_j,
    \]

  where:
  \begin{itemize}
    \item \(P_j\) is a continuous variable for weeks since discontinuation \(j\)
    \item \(T\) is the treatment arm indicator, which is coded 0 for the control treatment arm and 1 for the active treatment arm.
\end{itemize}
    
    We used 100 imputations, and ANCOVA results from each imputed dataset were combined using Rubin's rules to obtain the overall estimate and inference. 
   
	\item \textbf{Reference-based imputation:} We applied reference-based imputation while accounting for the availability of post-ICE data. Specifically, we used the reference-based imputation method corresponding to the true value of \(k_0\): J2R was applied when the true \(k_0 = 0\), and CIR was applied when the true \(k_0 = 1\).
    Unlike the retrieved-dropout approach, the imputation model fit does not use post-ICE outcomes in the imputation model; instead, only pre-ICE data contribute to the fitting of the imputation model. In our setting, where each patient either has all post-ICE values observed or all missing, any available post-ICE data are incorporated at the analysis stage. This method was implemented using the \texttt{rbmi} R package \cite{RN78}.

    The Bayesian models were fitted using 200 burn-in iterations, and posterior draws were thinned by retaining every 50th iteration. A total of 100 imputations were generated to obtain estimates under the J2R and CIR reference-based assumptions, with final inference conducted using Rubin’s rules. Convergence diagnostics of model fits were examined in an exploratory fashion for 50 simulation runs and indicated satisfactory mixing of chains, with effective sample sizes exceeding 1000 for all parameters.

	\item \textbf{Bayesian causal model:} The BCM was implemented under the Bayesian framework using Stan by specifying separate likelihood contributions for the pre- and post-ICE data in the active treatment arm. The pre-ICE data likelihood contributions were specified using the MMRM model, and likelihood contributions corresponding to the conditional density of the post-ICE data conditional on the pre-ICE data were specified using the conditional distribution in Equation \ref {eq:1_paper2} as described in Section 3. The treatment effect at the final time point was then estimated using the posterior means of $\hat \theta_{CB_1}$ obtained using Equation \ref{eq:2_paper2}. We used non-informative normal priors for the $k_0$ parameters, i.e., $N(0,10000)$, which allows posterior inference to be driven primarily by the data. We conducted further analyses with informative priors for $k_0$ centred at the true values of $k_0$, specifically $N(0,0.25)$ and $N(1,0.25)$. We applied the same priors under the BCM imputation method.
    
    A Dirichlet prior was assumed for the proportions of treatment discontinuation at each visit in the active treatment arm. We specified non-informative normal priors \(N(0,100)\) for the \(\mu\) parameters. The covariance matrix for $\Sigma$ was assigned weakly informative priors via an LKJ prior on the correlation structure, with non-informative priors on the marginal standard deviations.

    We performed model diagnostics to assess the performance of the Stan models using 50 simulations in an exploratory fashion, and these indicated good mixing of the chains. The effective sample size was sufficient to support reliable posterior inference for all the parameters. Based on the diagnostics, we selected a burn-in of 300 iterations and retained 1,000 post–warm-up iterations to obtain posterior samples under the BCM.
	
	\item \textbf{BCM with conditional mean imputation plus jackknife and bootstrap:} In the same way as for the multiple imputation approach, MAP estimates for the causal imputation model in Equation~\ref{eq:1_paper2} were obtained using \texttt{rstan::optimizing()} for each jackknife/bootstrap sample. The same priors as those used in the multiple imputation approach were applied. These MAP estimates were then used to impute missing post-ICE outcomes via their conditional means, and the imputed data were again analysed by ANCOVA. Standard errors (SEs) for the treatment effect were calculated using both the jackknife and bootstrap methods using 200 bootstrap samples.

     \item \textbf{BCM with multiple imputation plus bootstrapping:}
    We implemented multiple imputation using a bootstrap-based procedure. For each of \(B=200\) bootstrap resamples, we created \(m=50\) imputed complete datasets, analysed each separately with ANCOVA, and combined the results to obtain the average treatment effect for that resample. The final treatment-effect estimate was the mean across the \(B\) resamples, and its standard error was computed from the bootstrap distribution. Within each bootstrap sample, the MAP estimates were obtained via \texttt{rstan::optimizing()} and used to generate the multiple imputations. 
    
\end{itemize}

Annotated code for the simulations is available at \href{https://github.com/brendahnansereko/Bayesian_causal_model_post_ICE.git}{this GitHub repository}(\url{https://github.com/brendahnansereko/Bayesian_causal_model_post_ICE.git})

\subsection{Simulation results}

Tables~\ref{simTableHighIcek0_0_paper2} ($k_0=0$) and~\ref{simTableHighIcek0_1_paper2} ($k_0=1$) summarise the estimated mean treatment effects at the final time point for the higher ICE scenario, along with the empirical standard error (Emp.SE), average model-based standard error (Est.SD), and coverage probability. Coverage is defined as the percentage of times the 95\% interval (Bayesian or otherwise) includes the true value of the parameter for the treatment effect in the data-generating mechanism. Tables~\ref {simTableLowIcek0_0_paper2} and ~\ref {simTableLowIcek0_1_paper2} show the results for the low ICE rate scenario.

In the complete data analyses, the Emp.SE and Est.SD were consistently lower under the BCM compared to the ANCOVA model at the final time point, particularly in the high ICE scenario. This reflects the fact that the Bayesian model exploits the occurrence of the ICE in the estimation of the treatment effect, even with complete data, in contrast to the ANCOVA model. 

Introducing 20\% missingness in post-ICE outcomes resulted in a modest increase in Emp.SE and Est.SD for the BCM and reference-based methods, relative to the complete data analysis. However, with 90\% missingness, the increase in Emp.SE and Est.SD was substantially larger for the RD method, whereas the rise under the BCM remained more modest. Similar to findings in previous studies, while Est.SD increased with higher missingness levels, Emp.SE decreased for the reference-based estimators with Rubin's rules.

In contrast to the reference-based methods using Rubin’s rules, which showed a larger Est.SD than the Emp.SE, the Emp.SE and Est.SD for both the BCM and the imputation BCM approaches were closely aligned under both the low and high ICE scenarios, with both 20\% and 90\% missing post-ICE data. Under lower ICE scenarios, the Emp.SEs and Est.SDs were similar for both BCM and imputation BCM under the 20\% and 90\% missing post-ICE data scenarios. In contrast, under higher ICE scenarios, both Emp.SE and Est.SDs were smaller for BCM than for imputation BCM, especially under the 20\% missing post-ICE data scenario. This is due to the fact that the BCM imputation approach is only using the BCM model for the imputation of missing data.

Under the high-ICE scenario, the Emp.SE and Est.SD for the fully Bayesian BCM with 20\% missing post-ICE data were lower than those from the complete-data ANCOVA model; however, under 90\% missing post-ICE data, both Emp.SE and Est.SD increased above those of the complete-data ANCOVA model. In contrast, under the imputation-based BCM, Emp.SE and Est.SD were higher than those from the complete-data ANCOVA model for the case of 90\% missing post-ICE data scenarios, but were approximately equal to those of ANCOVA when only 20\% of post-ICE data were missing.
Across both the fully Bayesian and imputation-based BCM approaches, Emp.SE and Est.SD were larger when a non-informative prior (\(\sigma_{k_0}=100\)) was used compared with a mildly informative prior (\(\sigma_{k_0}=0.5\)), with the difference most pronounced under the high-ICE scenario with 90\% missing post-ICE data, as one would expect.

Under both high and low ICE scenarios with low missingness, the BCM, the imputation BCM and the retrieved dropout methods achieved coverage probabilities close to the nominal 95\%. In contrast, with high missingness, RD methods exhibited poor coverage performance. The BCM and the imputation BCM maintained good coverage with 95\% under high rates of missing post-ICE data. On the other hand, reference-based imputation methods using Rubin’s rules has coverage exceeding 95\%—when the assumed post-ICE data generation mechanism (e.g., J2R or CIR) was correct. 

The bias in the estimated treatment effect was minimal for most methods under low missingness, but varied substantially under high missingness, particularly for reference-based approaches. The BCM, the imputation BCM and the RD methods consistently produced estimates close (on average) to the true values across both high and low ICE scenarios, even with up to 90\% missingness.  Reference-based methods with Rubin's rules exhibited greater sensitivity to high missingness, with mean estimates deviating more markedly from the true effect. 

The standard errors under the BCM imputation method were estimated using either the jackknife or bootstrap methods. The Est.SD were similar under both the low-ICE and high-ICE scenarios with 20\% missingness; however, under 90\% missingness, the jackknife estimates were slightly higher than those obtained via bootstrapping, as expected, since the jackknife method tends to overestimate standard errors. The only method that had good coverage, little bias, and SEs larger than the complete method in all scenarios was the BCM imputation method.

\begin{table}[H]
\centering
\scriptsize
\caption{Estimated treatment effect at the final visit under the higher ICE rate scenario with $N=500$ (True $k_0=0$), true treatment effect=-0.388}
\label{simTableHighIcek0_0_paper2}
\resizebox{\textwidth}{!}{%
\begin{tabular}{@{} l c c
  S[table-format=-1.3]
  S[table-format=1.3]
  S[table-format=1.3]
  S[table-format=2.1]
@{}}
\toprule
Method & $k_0$ prior SD & \% missing post ICE & {Mean} & {Emp.SE} & {Est.SE} & {Cov} \\
\midrule
{\textbf{Complete data analysis}} \\
BCM & $\sigma_{k_0}=100$ & 0  & -0.383 & 0.068 & 0.065 & 93.6 \\
ANCOVA & --- & 0  & -0.389 & 0.078 & 0.079 & 95.4 \\
\midrule

\multirow{4}{*}{\textbf{BCM}}
& \multirow{2}{*}{$\sigma_{k_0}=100$} & 20 & -0.383 & 0.071 & 0.067 & 93.6 \\
&                                   & 90 & -0.380 & 0.104 & 0.101 & 94.2 \\
\cline{2-7}
& \multirow{2}{*}{$\sigma_{k_0}=0.5$}  & 20 & -0.385 & 0.068 & 0.066 & 94.0 \\
&                                    & 90 & -0.384 & 0.085 & 0.090 & 96.2 \\
\midrule

\multirow{6}{*}{\textbf{BCM Conditional mean imputation}}
& \multirow{2}{*}{$\sigma_{k_0}=100$ -- JK} & 20 & -0.387 & 0.078 & 0.079 & 95.3 \\
&                                         & 90 & -0.390 & 0.102 & 0.109 & 96.4 \\
\cline{2-7}
& \multirow{2}{*}{$\sigma_{k_0}=100$ -- BS} & 20 & -0.388 & 0.079 & 0.078 & 94.8 \\
&                                         & 90 & -0.389 & 0.103 & 0.103 & 94.6 \\
\cline{2-7}
& \multirow{2}{*}{$\sigma_{k_0}=0.5$ -- BS}  & 20 & -0.388 & 0.078 & 0.078 & 94.8 \\
&                                         & 90 & -0.390 & 0.089 & 0.088 & 94.8 \\
\midrule

\multirow{4}{*}{\textbf{BCM Multiple Imputation}}
& \multirow{2}{*}{$\sigma_{k_0}=100$ -- BS} & 20 & -0.388 & 0.079 & 0.078 & 95.2 \\
&                                         & 90 & -0.390 & 0.102 & 0.104 & 94.8 \\
\cline{2-7}
& \multirow{2}{*}{$\sigma_{k_0}=0.5$ -- BS}  & 20 & -0.388 & 0.078 & 0.078 & 95.1 \\
&                                         & 90 & -0.390 & 0.088 & 0.088 & 94.8 \\
\midrule

\multirow{2}{*}{\textbf{RD imputation}}
& \multirow{2}{*}{---} & 20 & -0.392 & 0.080 & 0.083 & 95.6 \\
&                     & 90 & -0.418 & 0.184 & 0.164 & 92.2 \\
\midrule

\multirow{2}{*}{\textbf{J2R (Rubin's rules)}}
& \multirow{2}{*}{---} & 20 & -0.393 & 0.075 & 0.082 & 96.6 \\
&                     & 90 & -0.410 & 0.065 & 0.093 & 99.3 \\
\bottomrule
\end{tabular}%
}
\begin{threeparttable}
\begin{tablenotes}
\scriptsize
\item[] Emp.SE = Empirical standard error; Est.SE = Model-based standard error; Cov = Coverage; RD = Retrieved-dropout; BCM = full Bayesian causal model. Simulations based on 5000 datasets. JK = Jackknife; BS = Bootstrap. \\
Monte Carlo standard error of the mean estimates is below 0.0015 in all scenarios.
\end{tablenotes}
\end{threeparttable}
\end{table}

\begin{table}[H]
\centering
\scriptsize
\caption{Estimated treatment effect at the final visit under the higher ICE rate scenario with $N=500$ (True $k_0=1$), true treatment effect=-0.628}
\label{simTableHighIcek0_1_paper2}
\resizebox{\textwidth}{!}{%
\begin{tabular}{@{} l c c
  S[table-format=-1.3]
  S[table-format=1.3]
  S[table-format=1.3]
  S[table-format=2.1]
@{}}
\toprule
Method & $k_0$ prior SD & \% missing post ICE & {Mean} & {Emp.SE} & {Est.SE} & {Cov} \\
\midrule
{\textbf{Complete data analysis}} \\
BCM & $\sigma_{k_0}=100$ & 0  & -0.623 & 0.064 & 0.063 & 94.7 \\
ANCOVA & --- & 0 & -0.627 & 0.075 & 0.075 & 95.1 \\
\midrule

\multirow{4}{*}{\textbf{BCM}}
& \multirow{2}{*}{$\sigma_{k_0}=100$} & 20 & -0.623 & 0.067 & 0.066 & 94.4 \\
&                                   & 90 & -0.619 & 0.100 & 0.100 & 94.6 \\
\cline{2-7}
& \multirow{2}{*}{$\sigma_{k_0}=0.5$} & 20 & -0.624 & 0.065 & 0.066 & 94.7 \\
&                                   & 90 & -0.625 & 0.086 & 0.091 & 96.3 \\
\midrule

\multirow{6}{*}{\textbf{BCM Conditional mean imputation}}
& \multirow{2}{*}{$\sigma_{k_0}=100$ -- JK} & 20 & -0.626 & 0.077 & 0.079 & 96.0 \\
&                                         & 90 & -0.628 & 0.100 & 0.114 & 96.9 \\
\cline{2-7}
& \multirow{2}{*}{$\sigma_{k_0}=100$ -- BS} & 20 & -0.628 & 0.075 & 0.075 & 94.8 \\
&                                         & 90 & -0.627 & 0.102 & 0.101 & 94.1 \\
\cline{2-7}
& \multirow{2}{*}{$\sigma_{k_0}=0.5$ -- BS} & 20 & -0.626 & 0.075 & 0.075 & 94.4 \\
&                                         & 90 & -0.628 & 0.090 & 0.088 & 94.3 \\
\midrule

\multirow{4}{*}{\textbf{BCM Multiple Imputation}}
& \multirow{2}{*}{$\sigma_{k_0}=100$ -- BS} & 20 & -0.388 & 0.079 & 0.078 & 95.2 \\
&                                         & 90 & -0.625 & 0.098 & 0.101 & 95.5 \\
\cline{2-7}
& \multirow{2}{*}{$\sigma_{k_0}=0.5$ -- BS} & 20 & -0.629 & 0.075 & 0.075 & 95.0 \\
&                                         & 90 & -0.626 & 0.086 & 0.088 & 95.1 \\
\midrule

\multirow{2}{*}{\textbf{RD imputation}}
& \multirow{2}{*}{---} & 20 & -0.630 & 0.080 & 0.080 & 94.2 \\
&                     & 90 & -0.624 & 0.182 & 0.165 & 92.7 \\
\midrule

\multirow{2}{*}{\textbf{CIR (Rubin's rules)}}
& \multirow{2}{*}{---} & 20 & -0.627 & 0.073 & 0.079 & 96.6 \\
&                     & 90 & -0.628 & 0.070 & 0.091 & 98.9 \\
\bottomrule
\end{tabular}%
}
\begin{threeparttable}
\begin{tablenotes}
\scriptsize
\item[] Emp.SE = Empirical standard error; Est.SE = Model-based standard error; Cov = Coverage; RD = Retrieved-dropout; BCM = full Bayesian causal model. Simulations based on 5000 datasets. JK = Jackknife; BS = Bootstrap. \\
Monte Carlo standard error of the mean estimates is below 0.0015 in all scenarios.
\end{tablenotes}
\end{threeparttable}
\end{table}

\begin{table}[H]
\centering
\scriptsize
\caption{Estimated treatment effect at the final visit under the lower ICE rate scenario with $N=500$ (True $k_0=0$), true treatment effect=-0.625}
\label{simTableLowIcek0_0_paper2}
\resizebox{\textwidth}{!}{%
\begin{tabular}{@{} l c c
  S[table-format=-1.3]
  S[table-format=1.3]
  S[table-format=1.3]
  S[table-format=2.1]
@{}}
\toprule
Method & $k_0$ prior SD & \% missing post ICE & {Mean} & {Emp.SE} & {Est.SE} & {Cov} \\
\midrule
{\textbf{Complete data analysis}} \\
BCM & $\sigma_{k_0}=100$ & 0 & -0.621 & 0.073 & 0.072 & 94.6 \\
ANCOVA & --- & 0 & -0.624 & 0.076 & 0.077 & 95.0 \\
\midrule

\multirow{4}{*}{\textbf{BCM}}
& \multirow{2}{*}{$\sigma_{k_0}=100$} & 20 & -0.620 & 0.074 & 0.073 & 95.4 \\
&                                   & 90 & -0.620 & 0.085 & 0.086 & 95.2 \\
\cline{2-7}
& \multirow{2}{*}{$\sigma_{k_0}=0.5$} & 20 & -0.622 & 0.072 & 0.072 & 95.4 \\
&                                   & 90 & -0.623 & 0.077 & 0.078 & 95.9 \\
\midrule

\multirow{6}{*}{\textbf{BCM Conditional mean imputation}}
& \multirow{2}{*}{$\sigma_{k_0}=100$ -- JK} & 20 & -0.625 & 0.076 & 0.077 & 95.2 \\
&                                         & 90 & -0.625 & 0.086 & 0.090 & 95.7 \\
\cline{2-7}
& \multirow{2}{*}{$\sigma_{k_0}=100$ -- BS} & 20 & -0.624 & 0.077 & 0.077 & 95.3 \\
&                                         & 90 & -0.623 & 0.087 & 0.087 & 94.8 \\
\cline{2-7}
& \multirow{2}{*}{$\sigma_{k_0}=0.5$ -- BS} & 20 & -0.624 & 0.076 & 0.077 & 95.3 \\
&                                         & 90 & -0.624 & 0.077 & 0.077 & 95.0 \\
\midrule

\multirow{2}{*}{\textbf{BCM Multiple Imputation}}
& \multirow{2}{*}{$\sigma_{k_0}=100$ -- BS} & 20 & -0.624 & 0.077 & 0.077 & 95.8 \\
&                                         & 90 & -0.624 & 0.087 & 0.087 & 95.0 \\
& \multirow{2}{*}{$\sigma_{k_0}=0.5$ -- BS} & 20 & -0.624 & 0.076 & 0.077 & 95.2 \\
&                                         & 90 & -0.626 & 0.077 & 0.077 & 95.2 \\
\midrule

\multirow{2}{*}{\textbf{RD}}
& \multirow{2}{*}{---} & 20 & -0.628 & 0.079 & 0.078 & 95.0 \\
&                     & 90 & -0.653 & 0.118 & 0.109 & 91.6 \\
\midrule

\multirow{2}{*}{\textbf{J2R (Rubin's rules)}}
& \multirow{2}{*}{---} & 20 & -0.626 & 0.077 & 0.078 & 95.5 \\
&                     & 90 & -0.633 & 0.073 & 0.082 & 97.1 \\
\bottomrule
\end{tabular}%
}
\begin{threeparttable}
\begin{tablenotes}
\scriptsize
\item[] Emp.SE = Empirical standard error; Est.SE = Model-based standard error; Cov = Coverage; RD = Retrieved-dropout; BCM = full Bayesian causal model. Simulations based on 5000 datasets. JK = Jackknife; BS = Bootstrap. \\
Monte Carlo standard error of the mean estimates is below 0.0015 in all scenarios.
\end{tablenotes}
\end{threeparttable}
\end{table}

\begin{table}[H]
\centering
\scriptsize
\caption{Estimated treatment effect at the final visit under the lower ICE rate scenario with $N=500$ (True $k_0=1$) , true treatment effect=-0.707}
\label{simTableLowIcek0_1_paper2}
\resizebox{\textwidth}{!}{%
\begin{tabular}{@{} l c c
  S[table-format=-1.3]
  S[table-format=1.3]
  S[table-format=1.3]
  S[table-format=2.1]
@{}}
\toprule
Method & $k_0$ prior SD & \% missing post ICE & {Mean} & {Emp.SE} & {Est.SE} & {Cov} \\
\midrule
{\textbf{Complete data analysis}} \\
BCM & $\sigma_{k_0}=100$ & 0 & -0.702 & 0.072 & 0.071 & 95.0 \\
ANCOVA & --- & 0 & -0.705 & 0.075 & 0.075 & 94.9 \\
\midrule

\multirow{4}{*}{\textbf{BCM}}
& \multirow{2}{*}{$\sigma_{k_0}=100$} & 20 & -0.702 & 0.075 & 0.072 & 95.0 \\
&                                   & 90 & -0.703 & 0.086 & 0.085 & 94.6 \\
\cline{2-7}
& \multirow{2}{*}{$\sigma_{k_0}=0.5$} & 20 & -0.703 & 0.072 & 0.072 & 95.1 \\
&                                   & 90 & -0.706 & 0.075 & 0.079 & 96.2 \\
\midrule

\multirow{6}{*}{\textbf{BCM Conditional mean imputation}}
& \multirow{2}{*}{$\sigma_{k_0}=100$ -- JK} & 20 & -0.706 & 0.075 & 0.074 & 94.7 \\
&                                         & 90 & -0.706 & 0.084 & 0.089 & 95.7 \\
\cline{2-7}
& \multirow{2}{*}{$\sigma_{k_0}=100$ -- BS} & 20 & -0.707 & 0.077 & 0.077 & 95.0 \\
&                                         & 90 & -0.706 & 0.085 & 0.086 & 95.6 \\
\cline{2-7}
& \multirow{2}{*}{$\sigma_{k_0}=0.5$ -- BS} & 20 & -0.707 & 0.075 & 0.075 & 94.8 \\
&                                         & 90 & -0.706 & 0.077 & 0.077 & 95.4 \\
\midrule

\multirow{4}{*}{\textbf{BCM Multiple Imputation}}
& \multirow{2}{*}{$\sigma_{k_0}=100$ -- BS} & 20 & -0.388 & 0.078 & 0.078 & 94.7 \\
&                                         & 90 & -0.708 & 0.086 & 0.086 & 95.0 \\
\cline{2-7}
& \multirow{2}{*}{$\sigma_{k_0}=0.5$ -- BS} & 20 & -0.706 & 0.074 & 0.075 & 94.9 \\
&                                         & 90 & -0.709 & 0.078 & 0.077 & 95.0 \\
\midrule

\multirow{2}{*}{\textbf{RD}}
& \multirow{2}{*}{---} & 20 & -0.706 & 0.075 & 0.076 & 95.0 \\
&                     & 90 & -0.711 & 0.115 & 0.109 & 93.3 \\
\midrule

\multirow{2}{*}{\textbf{CIR (Rubin's rules)}}
& \multirow{2}{*}{---} & 20 & -0.705 & 0.075 & 0.076 & 95.3 \\
&                     & 90 & -0.706 & 0.074 & 0.080 & 97.1 \\
\bottomrule
\end{tabular}%
}
\begin{threeparttable}
\begin{tablenotes}
\scriptsize
\item[] Emp.SE = Empirical standard error; Est.SE = Model-based standard error; Cov = Coverage; RD = Retrieved-dropout; BCM = full Bayesian causal model. Simulations based on 5000 datasets. JK = Jackknife; BS = Bootstrap. \\
Monte Carlo standard error of the mean estimates is below 0.0015 in all scenarios.
\end{tablenotes}
\end{threeparttable}
\end{table}

\section{Application}

The proposed Bayesian causal model was applied to a publicly available dataset from a clinical trial evaluating the efficacy of duloxetine in improving emotional and painful physical symptoms in patients with depression \cite{RN55}. This data is available at \url{https://www.lshtm.ac.uk/research/centres-projects-groups/missingdata#dia-missing-data}. Participants were assessed at baseline and at weeks 1, 2, 4, 6, and 8 using the 17-item Hamilton Depression Rating Scale (HAMD17), with the primary outcome defined as the change in HAMD17 score from baseline. For the purposes of this analysis, data collected up to week 6 were used, with week 6 designated as the final visit. The trial enrolled a total of 171 participants, with 84 (48.5\%) randomised to the active treatment (duloxetine) arm and 88 (51.5\%) to the placebo arm. Among those in the active treatment group, 63 (24.0\%) experienced an ICE, while 65 (26.1\%) in the placebo arm experienced an ICE. Notably, participants were not followed after the ICE in the original trial. However, O'Kelly and Li, as part of the work of the Scientific Working Group on Estimands and Missing Data of the Drug Information Association (SWGEMD), simulated some post-ICE data under several assumptions post-ICE data \cite{RN94}. Two post-ICE datasets were generated based on post-ICE patterns defined by the timing of the ICE. In the {\em covered} dataset, all post-ICE patterns contained at least some observed post-ICE data. By contrast, the {\em perforated} dataset included one  post-ICE pattern for which there was no observed post-ICE data, such that all post-ICE outcomes were missing for subjects experiencing the ICE at that time point, as indicated in Table \ref{tab:coveredperf}.

\begin{table}[H]
	\centering
	\scriptsize
	\caption{Summary of pre-ICE, observed post-ICE, and missing post-ICE frequencies by visit for both the covered and perforated dataset}
	\label{tab:on_off_summary_paper2}
	\begin{tabular*}{\linewidth}{@{\extracolsep{\fill}}c|ccc|ccc|ccc|ccc}
		\hline
		\multirow{4}{*}{Visit} 
		& \multicolumn{6}{c|}{\textbf{Covered dataset}} 
		& \multicolumn{6}{c}{\textbf{Perforated dataset}} \\
		
		& \multicolumn{3}{c|}{\textbf{Control arm--n}} 
		& \multicolumn{3}{c|}{\textbf{Treatment arm--n}}
		& \multicolumn{3}{c|}{\textbf{Control arm--n}} 
		& \multicolumn{3}{c}{\textbf{Treatment arm--n}} \\
		
		& \multicolumn{1}{c}{\textbf{Pre-ICE}} & \multicolumn{2}{c|}{\textbf{Post-ICE}} 
		& \multicolumn{1}{c}{\textbf{Pre-ICE}} & \multicolumn{2}{c|}{\textbf{Post-ICE}} 
		& \multicolumn{1}{c}{\textbf{Pre-ICE}} & \multicolumn{2}{c|}{\textbf{Post-ICE}} 
		& \multicolumn{1}{c}{\textbf{Pre-ICE}} & \multicolumn{2}{c}{\textbf{Post-ICE}} \\
		
		\cline{2-13}
		& Pre & Obs & Miss & Pre & Obs & Miss & Pre & Obs & Miss & Pre & Obs & Miss \\
		\hline
		1 & 88 & 0 & 0 & 84 & 0 & 0 & 88 & 0 & 0 & 84 & 0 & 0 \\
		2 & 81 & 3 & 4 & 78 & 4 & 2 & 81 & 5 & 2 & 78 & 0 & 6 \\
		3 & 76 & 7 & 5 & 73 & 5 & 6 & 76 & 9 & 3 & 73 & 4 & 7 \\
		4 & 65 & 12 & 11 & 64 & 10 & 10 & 65 & 12 & 11 & 64 & 10 & 10 \\
		\hline
	\end{tabular*}
    \label{tab:coveredperf}
\end{table}

\begin{table}[H] 
\centering
\caption{Estimates of the treatment policy effect at the final time point (active vs placebo arm) in the antidepressant trial for the covered and perforated datasets.}
\label{tab:estimates_paper2}
\resizebox{\textwidth}{!}{%
\begin{tabular}{lll|cc|cc}
\hline
 &  &  & \multicolumn{2}{c|}{\textbf{Covered dataset}} 
 & \multicolumn{2}{c}{\textbf{Perforated dataset}} \\
\textbf{Model} & \textbf{Prior} &  & \textbf{Estimate} & \textbf{Est. SE}
               & \textbf{Estimate} & \textbf{Est. SE} \\
\hline
\multirow{3}{*}{BCM $k_0 \sim N(0,\sigma_{k_0}^2)$} 
& $\sigma_{k_0} = 0.1$   &  & -1.960 & 0.717 & -1.806 & 0.693 \\
& $\sigma_{k_0} = 1$     &  & -2.233 & 0.824 & -2.170 & 0.849 \\
& $\sigma_{k_0} = 100$   &  & -2.265 & 0.841 & -2.203 & 0.863 \\
\hline
\multirow{3}{*}{BCM $k_0 \sim N(1,\sigma_{k_0}^2)$} 
& $\sigma_{k_0} = 0.1$   &  & -2.470 & 0.925 & -2.452 & 0.926 \\
& $\sigma_{k_0} = 1$     &  & -2.287 & 0.857 & -2.218 & 0.862 \\
& $\sigma_{k_0} = 100$   &  & -2.265 & 0.831 & -2.190 & 0.858 \\
\hline
\multirow{3}{*}{BCM Imputation $k_0 \sim N(0,\sigma_{k_0}^2)$*} 
& $\sigma_{k_0} = 0.1$   &  & -2.316 & 0.919 & -2.404 & 0.927 \\
& $\sigma_{k_0} = 1$     &  & -2.324 & 0.978 & -2.379 & 0.961\\
& $\sigma_{k_0} = 100$   &  & -2.369 & 0.975 & -2.439 & 0.928 \\
\hline
\multirow{3}{*}{BCM Imputation $k_0 \sim N(1,\sigma_{k_0}^2)$*} 
& $\sigma_{k_0} = 0.1$   &  & -2.473 & 0.967 & -2.435 & 0.979 \\
& $\sigma_{k_0} = 1$     &  & -2.456 & 0.964 & -2.443 & 0.983 \\
& $\sigma_{k_0} = 100$   &  & -2.422 & 0.993 & -2.436 & 0.937 \\
\hline
\multirow{2}{*}{J2R-Retrieved-Reference} 
& $\sigma_{\gamma^*} = 1$    &  & -2.280 & 1.050 & -2.380 & 1.040 \\
& $\sigma_{\gamma^*} = 31.6$ &  & -2.320 & 1.100 & -2.630 & 2.800 \\
\hline
\multirow{2}{*}{CIR-Retrieved-Reference} 
& $\sigma_{\gamma^*} = 1$    &  & -2.410 & 1.040 & -2.420 & 1.040 \\
& $\sigma_{\gamma^*} = 31.6$ &  & -2.320 & 1.100 &-2.630 & 2.790 \\
\hline
\multicolumn{2}{l}{Retrieved-Dropout**}   &  & -2.234 & 1.112 & -- & -- \\
\multicolumn{2}{l}{J2R-Reference-based} &  & -2.238 & 1.050 & -2.441 & 1.035 \\
\multicolumn{2}{l}{CIR-Reference-based} &  & -2.429 & 1.040 & -2.457 & 1.031 \\
\hline
\end{tabular}
}
\begin{tablenotes}
\item[] *BCM conditional mean imputation approach with bootstrap standard errors. \\
        **Outcomes of some patient visits cannot be imputed due to non-estimable imputation parameters under the perforated dataset.
\end{tablenotes}
\end{table}

Table \ref{tab:estimates_paper2} presents estimates of the treatment policy effect from the antidepressant trial, based on the antidepressant covered  and perforated datasets. Multiple methods were employed to estimate this effect: the BCM, BCM conditional mean imputation method, traditional reference-based imputation approaches, and the RD imputation method (as described previously). We discarded the first 300 iterations as burn-in and used the remaining 10,000 iterations to estimate the posterior mean and standard deviation of the treatment effect under the BCM method. We applied 100 imputations for the RD and reference-based imputation methods and used 1,000 bootstrap samples for inference under the BCM conditional mean imputation method. For comparison, published results from the retrieved-reference approach proposed by Cro \textit{et al} are included (Table \ref{tab:estimates_paper2})  \cite{RN76}.

The BCM approach produced slightly smaller standard errors than the other methods, particularly when the prior mean for the $k_0$ parameter was set to $0$ and a smaller prior variance was used. 
Both the magnitude and precision of the treatment effect estimates varied depending on the prior assumptions specified for the BCM. For example, the estimated treatment effect decreased in magnitude to $-1.960$ under a tight prior with SD $\sigma_{k_0} = 0.1$, compared with an estimate of $-2.265$ under a diffuse prior with $\sigma_{k_0} = 100$, while keeping the same prior mean $\mu_{k_0} = 0$. Changes in the prior SD $\sigma_{k_0}$ had relatively small impacts on the point estimates and standard errors of the treatment effect under the BCM conditional mean imputation approach.  Overall, standard errors were lower under the BCM method compared with all the BCM imputation standard errors, in line with our earlier simulation results.

The point estimates obtained for the BCM conditional mean imputation were comparable to those from the reference--retrieved dropout, traditional reference-based imputation, and retrieved-dropout approaches. However, the standard errors were slightly smaller under the BCM conditional mean imputation compared with the reference--retrieved dropout approach under the covered dataset. In contrast, unlike the BCM conditional mean imputation approach, increasing the prior variances for the $\gamma_{akj}$ parameters in the reference-retrieved dropout approach led to a substantial increase in the standard errors under the perforated dataset. This behaviour is attributable to challenges in estimating the separate $\gamma_{akj}$ parameters for ICE patterns in which no post-ICE data were observed, whereas for the BCM, there is only a single parameter ($k_0$) to estimate.

We were unable to obtain an important result for the RD method under the perforated dataset. This was due to failures in imputing some missing post-ICE outcomes, arising from non-estimable imputation parameters. This highlights limitations of the RD method when post-ICE data are sparse. 

\section{Discussion}

Retrieved-dropout imputation is one of the most commonly used estimation approaches for handling ICEs under the treatment policy strategy. In this approach, observed data following an ICE are used to impute missing post-ICE outcomes. However, retrieved-dropout methods often suffer from variance inflation and model-fitting difficulties, particularly when post-ICE data are sparse, as seen in the perforated antidepressant dataset analysis. Traditional RBI methods are also employed to estimate treatment effects under the treatment policy strategy. Yet, these methods rely on strong assumptions about the distribution of post-ICE outcomes, which, if misspecified, can lead to biased estimates. To address these limitations, we build on White \textit{et al.}'s causal model within a Bayesian framework using the BCM, which provides inferences that reflect uncertainty in the reference-based assumptions while maintaining good frequentist properties, even when limited post-ICE data are available.

In our previous work, we introduced the BCM, which incorporated a prior on the maintained effect parameter ($k_0$) after treatment  and explored how choices of this prior affected inferences from this approach. That formulation assumed that no post-ICE data were available. However, the recent ICH E9(R1) guidelines emphasize the importance of collecting post-ICE data once a patient experiences an ICE. Motivated by this, in the current paper we have extended the BCM to explicitly incorporate post-ICE observations. The benefit of using the post-ICE data is to learn about $k_0$ from the available post-ICE data, which can improve imputation of missing post-ICE values and hence the estimation of the treatment effect. We have implemented this extension in two ways: (i) as a full Bayesian model, and (ii) through a BCM-based imputation framework. The latter employs conditional mean imputation and multiple imputation based on MAP estimates, with inferences obtained via the jackknife or bootstrap. 

Unlike the traditional reference-based methods, which effectively put a point prior on the proportion of treatment effect maintained after the ICE occurs, the proposed BCM allows for uncertainty in this parameter. Under high rates of missingness, the traditional reference-based methods with Rubin's rules yield a higher Est.SE compared to the Emp.SE. In contrast, under the BCM, the Est.SE and Emp.SE were in our simulations approximately equal, with frequentist coverage close to 95\%. Furthermore, while the Emp.SE decreases with increasing missingness under the traditional reference-based imputation methods (and may fall below what it would be if complete data were actually observed from patients under this assumption), in our simulations it increased with higher levels of missingness under the BCM. A key distinction is that traditional reference-based methods do not incorporate post-ICE data during the imputation model fitting process, whereas the BCM makes use of this information, which is likely to improve the accuracy of estimation and inference for the treatment effect in the presence of ICEs. 

In our simulations the retrieved dropout imputation method yielded higher Est.SE compared to the BCM, particularly when post-ICE data were sparse, which is commonly the case in practice. A moderate increase in Est.SE was observed as the amount of available post-ICE data decreased under the Bayesian and imputation BCM approach. An important advantage of the BCM approaches is that, when very little post-ICE data are available, the increase in Est.SE can be controlled through the prior variance of the $k_0$ parameter by specifying a more or less informative (strict) prior.

The BCM was implemented within an imputation framework, using both multiple and single conditional mean imputation combined with jackknife or bootstrap standard errors. Although our approach combined multiple imputation using the BCM with bootstrap or jackknife-based inference, Rubin’s rules could alternatively be used for inference, although we would not expect the resulting frequentist coverage to be correct. Under the low missingness scenario in our simulations, the imputation approaches yielded empirical and model-based standard errors similar to those obtained from the complete-data regression model. As the proportion of missing data increased, both standard errors increased moderately, reflecting the additional uncertainty due to missingness. These standard errors were larger than those from the full Bayesian BCM (without imputation), which is more precise because the treatment effect estimation utilises information on ICE occurrence. In contrast, the imputation-based BCM approach only utilises the BCM model at the imputation stage, leading to increased variability.

The BCM imputation framework may be particularly useful in settings with multiple types of ICEs, where different imputation strategies may be required for each type of ICE. Although multiple imputation is the most commonly used imputation approach, the single conditional mean imputation method does not rely on random sampling and therefore eliminates Monte Carlo error for the point estimate. Moreover, when combined with the jackknife, it also eliminates Monte Carlo error from the standard error estimate. However, in some settings, having multiple imputations drawn from the missing data distribution may be preferable, such as where the final analysis involves dichotomisation of the continuous outcome. Given these known advantages of MI in some situations, we implemented the MI approach with bootstrap standard errors, which yielded results similar to those obtained with conditional mean imputation and either bootstrap or jackknife standard errors.

From our simulations, we observed that the BCM imputation method exhibited good coverage and little bias. Under non-tight priors, the standard errors obtained using this method were larger than those from the complete-data analysis, which would usually be viewed as desirable and contrasts with the full Bayesian BCM, under which standard errors were smaller than those from the complete-data analysis. This suggests that the BCM imputation method may be preferable, since it only relies on the BCM model insofar as it is needed to impute missing data. When some post-ICE data are available, we believe the BCM imputation method is attractive, as it yields stable estimates and valid inference under mildly informative priors, provided the BCM modelling assumptions hold. In addition, the BCM imputation method requires substantially less computation time when implemented via the Stan optimisation function. We believe the BCM  imputation method should be considered as a viable potential approach for a trial's primary analysis, in line with recent FDA guidance on the use of Bayesian methodology in clinical trials of drug and biological products \cite{RN95}. In this context, it would be of crucial importance to agree on what is a reasonable choice for the prior on the maintained effect parameter $k_0$.

In our analyses of the antidepressant trial data, we observed smaller standard errors for the full Bayesian BCM compared with the imputation-based BCM and retrieved-reference methods across both the covered and perforated  datasets. The BCM imputation approaches yielded slightly smaller standard errors than the retrieved-reference methods in the covered dataset. However, in the perforated dataset, the standard errors for the retrieved-reference methods were substantially larger, particularly when using very wide prior variances ($\sigma_{\gamma})$. This is likely due to the increased difficulty of estimating multiple parameters across visits in the retrieved-reference approach, in contrast to the BCM framework, which requires estimation of only a single parameter $k_0$ across all visits.

In general, RBI and BCM methods are directly applicable to settings involving a single ICE corresponding to treatment discontinuation and a placebo control. However, the causal model underpinning the BCM may be extendable to more complex settings (e.g., those involving rescue medication), representing an important area for future research. However, this will likely come at the cost of an increased SE.  The causal model currently employed under the BCM may be somewhat restrictive, as it assumes that the maintained treatment effect parameter $k_0$ acts immediately after the ICE and remains constant thereafter. In many settings such as weight-loss trials, it may be more plausible to allow the maintained treatment effect to diminish gradually over time, and we are currently investigating such extensions.
Our method was developed under the assumption that the only missing data occur post-ICE and that, for a given participant, either all or none of the post-ICE data are observed. However, the approach could be extended directly to more general missing data patterns.


\section{Acknowledgments}
The authors gratefully acknowledge \textit{the UCL, Bloomsbury and East London Doctoral Training Partnership (UBEL DTP) and Roche for financial support for Nansereko's PhD studentship}. Jonathan Bartlett is supported by ESRC grant UKRI1720. James Carpenter is supported by MRC grant MC\_UU\_00004/07.

\section{Declaration of interest statement}
Jonathan Bartlett's past and present institutions have received consultancy fees for his advice on statistical methodology from AstraZeneca, Bayer, Novartis, and Roche. JB has in the past received consultancy fees from Bayer and Roche for statistical methodology advice.

\bibliographystyle{unsrtnat}

\bibliography{references.bib}

\end{document}